# Nonradiating toroidal structures

A D. Boardman[a] and K. Marinov[a*]
[a] Photonics and Nonlinear Science Group, Joule Laboratory, Department of Physics,
University of Salford, Salford M5 4WT, UK

N. Zheludev[b] and V. A. Fedotov[b]
[b] EPSRC Nanophotonics Portfolio Centre, School of Physics and Astronomy,
University of Southampton, Highfield, Southampton, SO17 1BJ, UK

**ABSTRACT**

Some basic properties of nonradiating systems are considered. A simple connection is established between the existence of residual electromagnetic potentials and the current density spectrum of the system. The properties of specific configurations based on toroidal and supertoroidal currents are modeled with the finite-difference time-domain method. Possible applications are discussed. A design of a new type of nonradiating system, based on a left-handed metamaterial is proposed and the system performance is modeled numerically.

**Keywords:** Nonradiating configurations, Toroids, Supertoroids, Finite-difference time-domain method, FDTD

## 1. INTRODUCTION

Nonradiating configurations are oscillating charge-current distributions that do not produce electromagnetic fields in the radiation zone. Such structures were initially considered in atomic physics [1, 2]. More recently, nonradiating configurations have been studied in connection with inverse-source problems [3-6] and the electrodynamics of toroidal and supertoroidal currents [7-9]. Nonradiating systems are of interest also in other branches of wave theory and physics (see e.g. [10, 11]).

A sufficient condition for absence of radiation from an arbitrary, localized charge distribution, exhibiting periodic motion with period $T = 2\pi/\omega_0$ is the absence of the Fourier components $\tilde{\mathbf{J}}(n\omega_0\mathbf{r}/cr, n\omega_0)$ from the spectrum of the current density $\mathbf{J}(\mathbf{r},t)$, where $c$ is the speed of light and $n$ is an integer number [1]. However, as pointed out [1], and later shown [12], this condition is not necessary. The results of [12] indicate that another condition, namely $\tilde{\mathbf{J}}(\omega\mathbf{r}/cr,\omega) \propto \mathbf{r}$, is a necessary and sufficient one. The latter condition only requires the absence of the Fourier components that are transverse to $\mathbf{r}$, as opposed to the absence of both the transverse and the longitudinal components [1]. It is interesting that the condition $\tilde{\mathbf{J}}(\omega\mathbf{r}/cr,\omega) \propto \mathbf{r}$ has appeared in an earlier work, [2], in connection with the self-oscillations of a non-relativistic particle.

An important conclusion that can be drawn from the earlier results is that two types of nonradiating systems can exist. The first type satisfies the condition $\tilde{\mathbf{J}}(\omega\mathbf{r}/cr,\omega) = 0$ and examples of such systems exist [1, 7]. A characteristic feature of these structures is that both the electromagnetic fields and the electromagnetic potentials are zero.

The Fourier spectrum of the second type of configurations is purely longitudinal i.e. $\tilde{\mathbf{J}}(\omega\mathbf{r}/cr,\omega) \propto \mathbf{r}$. Here the electromagnetic fields are zero but electromagnetic potentials may be finite.

It is pointed out [1] that the case $\tilde{\mathbf{J}}(\omega\mathbf{r}/cr,\omega) \propto \mathbf{r}$ corresponds to simple spherically symmetric oscillations of the charge density. However, non-trivial systems that satisfy this can be created using toroidal structures. It has been shown [7, 8] that a non-radiating configuration can be constructed by combining infinitesimal toroidal or supertoroidal solenoids with electric or magnetic dipoles placed in their center. The calculations performed in [7] and [8] show that while the electromagnetic fields disappear outside such a composite object, the electromagnetic potentials survive in certain cases. As we show here systems producing electromagnetic potentials in the absence of electromagnetic fields are those that satisfy $\tilde{J}(\omega r/cr,\omega) \propto r$.

---

[*] Further author information: (Send correspondence to K. M.)
A. D. B. E-mail: a.d.boardman@salford.ac.uk
K. M.E-mail: k.marinov@salford.ac.uk
N. Z. E-mail: n.i.zheludev@soton.ac.uk
V. A. F. E-mail: vaf@phys.soton.ac.uk

An important question concerning the non-radiating systems is what kind of applications these systems might have. The absence of radiation is, in fact, ensured by a specific relationship between the parameters of the system and those of the ambient environment. Variation of the value of any of these parameters will cause the system to radiate. In principle system that does not radiate under certain conditions could be employed to measure the parameters of the ambient environment. The present work addresses this question.

## 2. CONDITIONS FOR ABSENCE OF RADIATION AND ABSENCE OF ELECTROMAGNETIC POTENTIALS

Following some guidelines, set out by earlier works [1, 12], the conditions ensuring the absence of radiation from a charge-current distribution are derived in this section. The difference between absence of an electromagnetic field and absence of electromagnetic potential will be emphasized.

Consider the vector potential

$$A = \frac{\mu_0}{4\pi} \int \frac{J(r',t - |r-r'|/c)}{|r-r'|} d^3r'. \tag{1}$$

In the radiation zone the standard approximation $|r-r'| \approx r - r.r'/r$ [13] can be used and (1) reduces to

$$A = \frac{\mu_0}{4\pi r} \int J(r', t - r/c + r.r'/cr) d^3r'. \tag{2}$$

The current density $J(r,t)$ can be expressed through its Fourier-transform $\tilde{J}(k,\omega)$

$$J(r,t) = \int \tilde{J}(k,\omega) \exp(-i\omega t + ik.r) d^3k\, d\omega. \tag{3}$$

In (3) $k$ and $\omega$ are independent variables. Substitution of (3) in (2) yields

$$A = \frac{\mu_0 (2\pi)^3}{4\pi r} \int \tilde{J}\left(\frac{\omega r_0}{c}, \omega\right) \exp(-i\omega t + i\omega r/c) d\omega, \tag{4}$$

where $r_0 = r/r$. Equation (4) shows that the components of the Fourier-spectrum, which can generate electromagnetic waves, are the components corresponding to $|k| = \omega/c$. The components $\tilde{J}(k,\omega)$, corresponding to wavevectors $k$ and frequencies $\omega$, not related to each other by the dispersion equation of the wave, do not contribute to radiation.

Using equation (4) and the Lorentz gauge condition, $\operatorname{div} A + \frac{1}{c^2}\frac{\partial U}{\partial t} = 0$, the scalar potential $U(r,t)$ can be expressed as

$$U = \frac{(2\pi)^3}{4\pi\varepsilon_0 cr} \int r_0.\tilde{J}\left(\frac{\omega r_0}{c}, \omega\right) \exp(-i\omega t + i\omega r/c) d\omega. \tag{5}$$

Equations (4) and (5) clearly show that the charge-current configuration considered would emit no electromagnetic energy if the condition

$$\tilde{J}\left(|k| = \frac{\omega}{c}, \omega\right) = 0 \tag{6}$$

is satisfied. Equation (6) is a sufficient condition for absence of radiation [1]. It is clear that systems satisfying (6) produce no electromagnetic fields and no electromagnetic potentials. It can easily be shown, however, that (6) is not a necessary condition. Assuming monochromatic time dependence reduces (4) to

$$A = \frac{\mu_0 (2\pi)^3}{4\pi r} \tilde{J}\left(\frac{\omega r_0}{c}, \omega\right) \exp(-i\omega t + i\omega r/c). \tag{7}$$

In the radiation zone, the electromagnetic fields can then be obtained from (7) using $\mathbf{H} = \nabla \times \mathbf{A}/\mu_0$ and $\mathbf{E} = i \nabla \times \mathbf{H}/\omega \varepsilon_0$. The result is

$$\mathbf{E} = i\sqrt{\frac{\mu_0}{\varepsilon_0}} \frac{\omega(2\pi)^3}{4\pi cr} \mathbf{r}_0 \times (\tilde{\mathbf{J}} \times \mathbf{r}_0) \exp(-i\omega t + i\omega r/c) \qquad (8)$$

and

$$\mathbf{H} = i\frac{\omega(2\pi)^3}{4\pi cr} (\mathbf{r}_0 \times \tilde{\mathbf{J}}) \exp(-i\omega t + i\omega r/c). \qquad (9)$$

Using (8) and (9) the time-averaged Poynting vector, $\langle \mathbf{S} \rangle = (1/2)\mathbf{E} \times \mathbf{H}^*$, can be expressed in the form

$$\langle \mathbf{S} \rangle \propto |\mathbf{r}_0 \times (\tilde{\mathbf{J}} \times \mathbf{r}_0)|^2 \mathbf{r}_0. \qquad (10)$$

The quantity $\mathbf{r}_0 \times (\tilde{\mathbf{J}} \times \mathbf{r}_0)$ is the radiation pattern of the system. It is immediately clear form (10) that the charge-current distribution considered will emit no electromagnetic energy if

$$\tilde{\mathbf{J}}_\perp \equiv \mathbf{r}_0 \times (\tilde{\mathbf{J}} \times \mathbf{r}_0) = 0 \qquad (11)$$

which is a weaker sufficient condition compared to (6). The fact that it is also a necessary condition can be seen by setting $\mathbf{E}$ and $\mathbf{H}$ to zero in (8) and (9) [12]. Using the identity $\tilde{\mathbf{J}} = \mathbf{r}_0 \times (\tilde{\mathbf{J}} \times \mathbf{r}_0) + \mathbf{r}_0 (\mathbf{r}_0 \cdot \tilde{\mathbf{J}}) = \tilde{\mathbf{J}}_\perp + \tilde{\mathbf{J}}_\parallel$, the vector potential, as given by (7), can be expressed as

$$\mathbf{A} = \frac{\mu_0 (2\pi)^3}{4\pi r} (\tilde{\mathbf{J}}_\perp + \tilde{\mathbf{J}}_\parallel) \exp(-i\omega t + i\omega r/c). \qquad (12)$$

From (6), (11) and (12) it is clear that for a system satisfying (6) both the electromagnetic fields and the electromagnetic potentials are zero. In contrast, a configuration satisfying the weaker condition (11) radiates no electromagnetic energy but the electromagnetic potentials are finite. These potentials are given by (12) and (5).

## 3. TOROIDAL AND SUPERTOROIDAL DIPOLE MOMENTS

The toroidal (dipole) moments are easy to see as a generalization of the usual magnetic (dipole) moment. A rigorous approach to the problem is given in [7-9, 14].

Consider a wire loop of radius $d$ carrying a current of strength $I$ (Fig. 1a). A magnetic dipole moment $\mathbf{m}$ can be assigned to this loop according to the definition [13]

$$\mathbf{m} \equiv \mathbf{T}_0 = \frac{1}{2} I \oint \mathbf{r} \times d\mathbf{r} = \frac{1}{2} \oint \mathbf{r} \times d\mathbf{I} = \pi d^2 I \mathbf{n}. \qquad (13)$$

In (13) $\mathbf{T}_0$ stands for toroidal moment of zeroth order and $\mathbf{n}$ is a vector of unit length perpendicular to the plane of the loop. The current loop can be seen as a circular arrangement of current elements $d\mathbf{I} = I d\mathbf{r}$. Now if each current element is formally replaced with a magnetic dipole moment, a circular arrangement of dipole moments, known as a first-order toroidal moment, will be obtained (Fig 1b). The corresponding definition is [14]

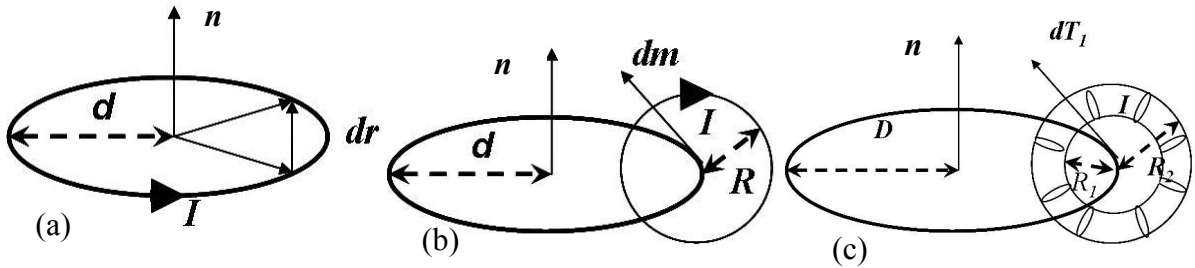

**Figure 1.** The first three members of the toroidal hierarchy: a current loop (a), toroidal solenoid (b) and a second-order supertoroid (c). One particular winding of the toroidal and the supertoroidal solenoids is shown in (b) and (c), respectively.

$$T_1 = \frac{1}{2}\oint r \times dm . \tag{14}$$

Replacing the current elements with magnetic dipole moments corresponds to constructing a toroidal solenoid with its windings wound along the meridians of a toroidal surface. The latter is shown in Fig. 2 in some more detail. Expressing *dm* through the total number of windings, $N$, and their radius $R$ according to

$$dm = \frac{N}{2\pi d}\pi R^2 I dr , \tag{15}$$

substituting this in (14), and using $\oint r \times dr = 2\pi d^2 n$ yields

$$T_1 = \frac{N\pi d R^2 I}{2} n . \tag{16}$$

Pursuing the same idea a step further, each of the windings of the toroidal solenoid can in turn be replaced with a toroidal solenoid (Fig. 1c). The resulting structure is a circular arrangement of (first-order) toroidal moments and it is known as second-order supertoroid [7-9]. The quantity characterizing this structure is the second-order toroidal dipole moment given by

$$T_2 = \frac{1}{2}\oint r \times dT_1 . \tag{17}$$

Note that *dI* in (13) has been replaced with *dm* to obtain (14), which in turn has been replaced by $dT_1$, to yield (17). Using the notation of Fig. 1c $dT_1$ can be expressed as

$$dT_1 = \frac{2\pi D}{R_2 - R_1}\frac{\pi N I (R_1 + R_2)(R_2 - R_1)^2}{16}\frac{dr}{2\pi D} . \tag{18}$$

In (18) $N$ is the number of windings in each first-order toroidal solenoid and the quantity $\frac{\pi N I (R_1 + R_2)(R_2 - R_1)^2}{16}$ is simply the toroidal moment of each of the first-order toroidal solenoids composing the second-order supertoroid. The factor $\frac{2\pi D}{R_2 - R_1}$ is the total number of first-order solenoids ($\frac{2\pi D}{R_2 - R_1}\frac{1}{2\pi D}$ is the number of first order solenoids per unit length). Thus (18) gives the $T_1$ - moment in an interval of length $|dr|$. Substituting (18) into (17) yields

$$T_2 = \frac{\pi^2 D^2 N I (R_2^2 - R_1^2)}{16} n . \tag{19}$$

The same procedure can be applied once again and each first-order toroidal solenoid can be replaced by a second-order one. This will result in a third-order supertoroid and so on.

## 4. NONRADIATING CONFIGURATIONS BASED ON INFINITESIMAL TOROIDAL AND SUPERTOROIDAL CURRENTS

The nonradiating configurations considered in this section are known from the literature [7, 8]. Here we propose a specific application for such systems, namely the possibility of measuring the permittivity of dielectric media. Then we give an interpretation of the ability of some configurations to generate finite time-dependent electromagnetic potentials in the absence of electromagnetic fields in terms of the general criteria for absence of radiation. These criteria have already been discussed in section 2.

Consider a first-order toroidal solenoid with $N$ windings carrying a current of strength $I$ Fig. 2. The current density can be expressed as

$$j_p = \nabla \times M \tag{20}$$

where $M = (0, M_\varphi, 0)$ is the magnetization vector and $M_\varphi$ is given by

$$M_\varphi = \begin{cases} \frac{NI(t)}{2\pi\rho}, & (\rho - d)^2 + z^2 \leq R^2 \\ 0, & (\rho - d)^2 + z^2 > R^2 . \end{cases} \tag{21}$$

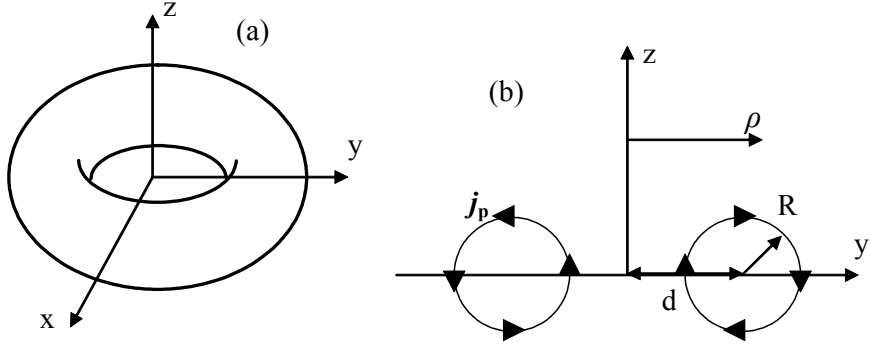

**Figure 2.** A toroidal surface $(\rho - d)^2 + z^2 = R^2$ in (a). The surface current $j_p$, flowing along the meridians of the toroid (poloidal current) is shown in (b).

The magnetization can in turn be expressed as

$$\mathbf{M} = \nabla \times \mathbf{T}, \tag{22}$$

where $\mathbf{T} = (0, 0, T_z)$ is the toroidization vector [7, 9]. The problem of calculating the electromagnetic fields radiated by a time-varying toroidal current is considerably simplified in the limit $d \to 0$. In this limit, the toroidization vector is

$$\mathbf{T} = T_1 \delta^3(\mathbf{r}), \tag{23}$$

where $T_1$ is given by (16). From a practical point of view (23) is equivalent to $d \ll \lambda$, where $\lambda$ is the radiation wavelength.

Assuming monochromatic time-dependence, $\propto \exp(-i\omega t)$, for the current flowing in the windings of the toroidal solenoid and using (20), (22) and (23), the magnetic field of the toroidal solenoid can be obtained in the form

$$\mathbf{H}_{T1} = \frac{k^2 T_1}{4\pi r^2}\left(\frac{1}{r} - ik\right)(\mathbf{n} \times \mathbf{r})\exp(ikr), \tag{24}$$

where $k$ is the wavenumber.

An electric dipole can now be introduced in the center of the toroidal solenoid. The current density associated with the composite object is

$$\mathbf{J}_1(\mathbf{r}) = T_1 \nabla \times \nabla \times \delta^{(3)}(\mathbf{r})\mathbf{n} + I_d L_d \delta^{(3)}(\mathbf{r})\mathbf{n}. \tag{25}$$

In (25) $L_d$ is the length of the dipole and $I_d$ is the amplitude of the dipole current. The electromagnetic field of the dipole is known [13].

The time-averaged power $P$, emitted by the structure, is

$$P = \frac{\mu_0 c k^2}{12\pi\sqrt{\varepsilon}}\left(I_d L_d + k^2 T_1\right)^2. \tag{26}$$

Equation (26) can be rewritten in the equivalent form

$$P = \frac{\mu_0 (\omega I_d L_d)^2}{12\pi c}\sqrt{\varepsilon}\left(1 - \frac{\varepsilon}{\tilde{\varepsilon}}\right)^2, \tag{27}$$

where

$$\tilde{\varepsilon} = -\frac{I_d L_d c^2}{\omega^2 T_1} \tag{28}$$

is the relative dielectric pemittivity of the medium in which the electromagnetic fields of the electric dipole and the first-order toroidal solenoid compensate each other. As (28) shows the value of $\tilde{\varepsilon}$ can be controlled by varying the ratio between the amplitudes of the currents flowing in the electrical dipole and the toroidal solenoid. This suggests that it should be possible to measure the relative dielectric permittivities of media (e.g. liquids) by adjusting experimentally the ratio of the currents flowing in the dipole and the toroidal solenoid until a zero of the emitted power is detected. The relative dielectric constant of the material under investigation can then be determined from (28).

To illustrate this point further, consider a second-order toroidal solenoid coaxial with a magnetic dipole [8]. The current density associated with this object is

$$\boldsymbol{J}_2 = T_2 \nabla \times \nabla \times \nabla \times \delta^{(3)}(\boldsymbol{r})\boldsymbol{n} + m \nabla \times \delta^{(3)}(\boldsymbol{r})\boldsymbol{n} \ . \tag{29}$$

The power emitted by this structure is

$$P = \frac{\mu_0 c k^4}{12\pi\sqrt{\varepsilon}} (m + k^2 T_2)^2 \ . \tag{30}$$

To see the dependence on the relative dielectric permittivity of the ambient material (30) can be rewritten as

$$P = \frac{\mu_0 \omega^4 m^2}{12\pi c^3} \varepsilon\sqrt{\varepsilon}\left(1 - \frac{\varepsilon}{\widetilde{\varepsilon}}\right)^2 \tag{31}$$

where

$$\widetilde{\varepsilon} = -\frac{mc^2}{\omega^2 T_2} \tag{32}$$

has the same physical meaning as in (28). By comparing (27) and (31) it is easy to see that the dependence of the latter on the dielectric permittivity of the ambient material is stronger. Indeed, for the structure consisting of a first-order toroidal solenoid and a electric dipole $P \propto \sqrt{\varepsilon}(1 - \varepsilon/\widetilde{\varepsilon})^2$, while for a combination of a second-order toroidal solenoid and a magnetic dipole $P \propto \varepsilon\sqrt{\varepsilon}(1 - \varepsilon/\widetilde{\varepsilon})^2$. This implies that a configuration involving higher-order toroidal solenoids may provide higher accuracy if used in dielectric permittivity measurements.

Let us now apply the criteria (6) and (11) to the two structures considered here. The current densities of the two systems are given by (25) and (29), respectively. Their Fourier-transforms are

$$(2\pi)^3 \widetilde{\boldsymbol{J}}_1(\boldsymbol{k}) = -T_1 \boldsymbol{k}(\boldsymbol{k}.\boldsymbol{n}) + (k^2 T_1 + I_d L_d)\boldsymbol{n} \tag{33}$$

and

$$(2\pi)^3 \widetilde{\boldsymbol{J}}_2(\boldsymbol{k}) = (k^2 T_2 + m)\boldsymbol{k} \times \boldsymbol{n} \ . \tag{34}$$

As discussed in Section 2 the electromagnetic radiation is generated by Fourier-components corresponding to wavenumbers $\boldsymbol{k} = \omega \boldsymbol{r}_0 \sqrt{\varepsilon}/c$. Therefore it is relevant to consider (33) and (34) only for $|\boldsymbol{k}| = \omega\sqrt{\varepsilon}/c$. By comparing (33) with (26) and (34) with (30) it is easy to see that in the absence of radiation the Fourier-spectra of the current densities of the two systems are given by

$$(2\pi)^3 \widetilde{\boldsymbol{J}}_1\left(\frac{\omega \boldsymbol{r}_0}{c}\right) = -T_1\left(\frac{\omega}{c}\right)^2 \varepsilon \boldsymbol{r}_0 (\boldsymbol{r}_0.\boldsymbol{n}) \tag{35}$$

and

$$\widetilde{\boldsymbol{J}}_2\left(\frac{\omega \boldsymbol{r}_0}{c}\right) = 0 \ . \tag{36}$$

As (36) shows, a combination of a magnetic dipole and a second-order toroidal solenoid satisfies the condition (6). As discussed in section 2, both the electromagnetic potentials and the electromagnetic fields are zero for such a system. According to (35) however, a nonradiating configuration built from a first-order toroidal solenoid and an electric dipole satisfies the weaker condition (11). Such systems produce finite electromagnetic potentials, in the absence of electromagnetic fields, due to the non-zero Fourier-component $\widetilde{\boldsymbol{J}}_\parallel$. The residual electromagnetic potentials associated with a combination of a first-order toroidal solenoid and an electric dipole are known from the literature [7, 8]. We have shown that *any* nonradiating system (and in particular this one) satisfying $\widetilde{\boldsymbol{J}}_\parallel(|\boldsymbol{k}| = \omega\sqrt{\varepsilon}/c) \neq 0$ has the *same* property.

# 5. NUMERICAL MODELING OF TOROIDAL RADIATORS

Although analytical formulae do exist for the electromagnetic fields of a toroidal solenoid [9], numerical evaluation of the latter is not straightforward beyond the limit $d \ll \lambda$. An approximations-free approach to the problem, based on an exact numerical solution of the Maxwell's equations is therefore useful, especially in the presence of inhomogeneities [15].

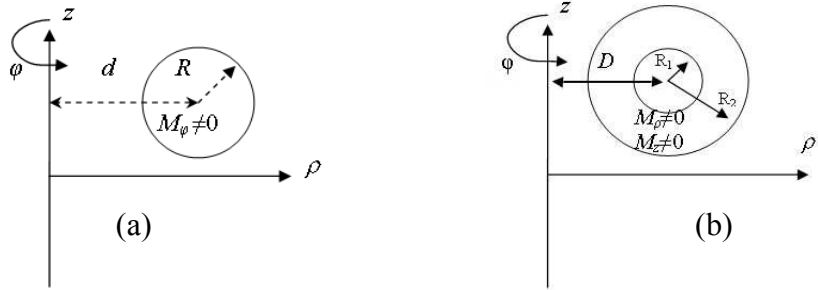

**Figure 3.** The cross-sections of toroidal solenoids with the $\rho$-$z$ plane – first order toroid in (a) and second-order in (b).

Since by definition the toroids are bodies of revolution (BOR), a special implementation of the finite-difference time-domain method (FDTD), known as BOR-FDTD [16] is well-suited to the problem. The method takes advantage of the azimuthal symmetry that reduces the problem to a two-dimensional one. Maxwell's equations are solved in the $\rho$-$z$ plane with the actual three-dimensional current distribution replaced by its cross-section with the $\rho$-$z$ plane (Fig 3). The three-dimensional distribution of the electromagnetic fields is then restored by rotating the two-dimensional solution around the $z$-axis.

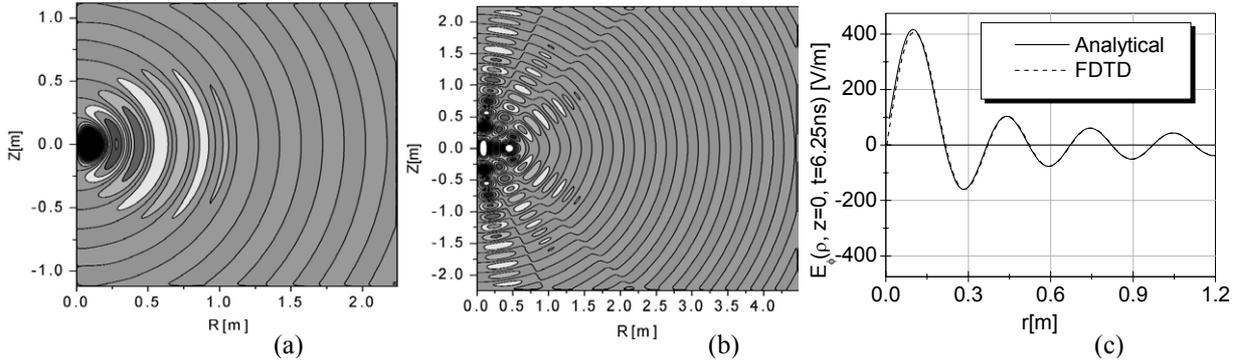

**Figure 4.** Electromagnetic radiation from different toroidal emitters driven by monochromatic currents at 1 GHz. Gray-scale map of the azimuthal magnetic field component $H_\varphi(\rho, z, t$=50ns) for a first-order toroidal solenoid with $d$=0.04m and $R$=0.02m in (a) and $d$=0.5m and $R$=0.3m in (b). Second order toroidal solenoid with $D$=0.08m, $R_2$=0.06m and $R_1$=0.04m in (c): comparison between the analytical and the numerical result for the azimuthal electric field component $E_\varphi(\rho, z$=0, $t$=6.25ns). The analytical result is obtained in the limit $D \ll \lambda$ with $\boldsymbol{T}_2$ given by (19).

For a first-order toroidal solenoid this cross-section is a circle, shown in Fig. 3a, confining the azimuthal magnetization (21). The cross-section of a second-order toroidal solenoid with the $\rho$-$z$ plane is a ring of magnetization, Fig 3b. The lines of the magnetization are circles coaxial with the ring. Away from the center the modulus of the magnetization vector decreases as the inverse of the radius of the circle. As Fig. 4a and 4c show the radiation patterns of small ($d \ll \lambda$) toroidal solenoids are dipole-like. In contrast, if their size is comparable with the wavelength (see Fig. 4b) the higher-order toroidal multipoles become important and the radiation pattern develops multiple peaks.

# 6. ON THE EQUIVALENCE OF THE RADIATION PATTERNS OF SUPERTOROIDS AND DIPOLES IN THE PRESENCE OF AN INTERFACE BETWEEN TWO DIELECTRIC MATERIALS

The fact that the radiation pattern of an infinitesimal supetoroidal solenoid is identical to that of a dipole in a homogeneous material is known (see e.g. [8], [14] and the references therein). In what follows we show that this is also the case in the presence of a planar interface separating two different dielectric materials.

Consider an infinitesimal second order toroidal solenoid coaxial with a magnetic dipole both located at the origin of a cylindrical system of coordinates (see Fig. 5). The dielectric permittivity of the material is equal to $\varepsilon$ for $z>d_0$ and to $\varepsilon_1$ for $z<d_0$.

The problem of a dipole radiating near the interface between two media with different dielectric properties was first investigated by A. Sommerfeld in 1909 [17]. Since then it has become a subject of a strong research interest (see e.g. [17-19]).

The electromagnetic wave in the region $z>0$ (Fig. 5) propagates in the positive $z$-direction. It has two components: a primary wave, originating at the source and a wave reflected by the interface. In the region $d_0<z<0$ the field contains an upward-propagating ($+z$) reflected component and a primary component propagating along $-z$. In the region $z<d_0$ the transmitted wave propagates along $-z$. The electromagnetic field is presented in the form of a superposition of cylindrical waves with identical radial wavenumbers in both the materials.

The solution is [19]

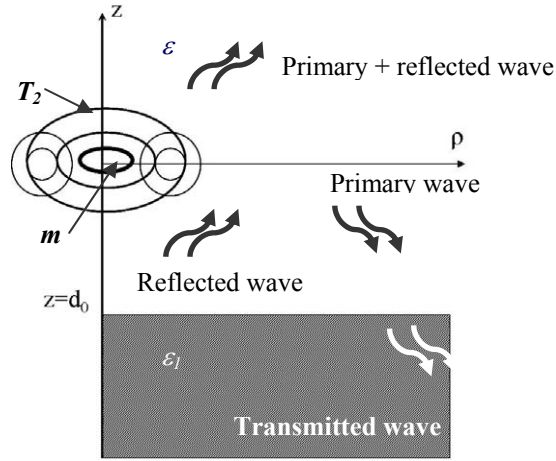

**Figure 5.** Second-order toroidal solenoid $T_2$ coaxial with a magnetic dipole $m$ radiating near the interface between two dielectric materials with relative dielectric permittivities $\varepsilon$ and $\varepsilon_1$.

$$H_z(\rho,z) = \int_{-\infty}^{\infty} H_s(k_\rho)\left[1 + R_{01}\exp(-2ik_{0z}d_0)\right]\exp(ik_{0z}z)H_0^{(1)}(k_\rho\rho)dk_\rho, \quad z>0 \tag{37a}$$

$$H_z(\rho,z) = \int_{-\infty}^{\infty} H_s(k_\rho)\left[R_{01}\exp(-2ik_{0z}d_0)\exp(ik_{0z}z) + \exp(-ik_{0z}z)\right]H_0^{(1)}(k_\rho\rho)dk_\rho, \quad d_0<z<0, \tag{37b}$$

$$H_z(\rho,z) = \int_{-\infty}^{\infty} H_s(k_\rho)[1+R_{01}]\exp(-ik_{0z}d_0)\exp(-ik_{1z}(z-d_0))H_0^{(1)}(k_\rho\rho)dk_\rho, \quad z<d_0. \tag{37c}$$

In (37) it is easy to identify the upward and downward propagating components of the electromagnetic field as well as the primary and the secondary (reflected) ones. The subscripts "0" and "1" refer to the media with dielectric permittivities $\varepsilon$ and $\varepsilon_1$, respectively. $H_0^{(1)}$ is the zeroth-order Hankel function of the first kind and $R_{01} = (k_{0z} - k_{1z})/(k_{0z} + k_{1z})$ is the Fresnel reflection coefficient for TE-waves. The axial wavenumbers are given by $k_{iz} = \sqrt{(\omega/c)^2 \varepsilon_i - k_\rho^2}$, where $i$=0, 1. The source function $H_S(k_\rho) = H_m(k_\rho) + H_{T2}(k_\rho)$ is the sum of the contributions of the dipole and the toroid. The dipole part of the source function [19] is $H_m(k_\rho) = \frac{im}{8\pi} \frac{k_\rho^3}{k_z}$.

Consider the axial magnetic field component of the supertoroid in a homogeneous material with dielectric permittivity $\varepsilon$

$$H_z = -k_0^2 T_2 \frac{1}{\rho} \frac{\partial}{\partial \rho} \rho \frac{\partial}{\partial \rho} \frac{\exp(ik_0 r)}{4\pi r}. \tag{38}$$

With the aid of the Sommerfeld's formula [19]

$$\frac{\exp(ikr)}{4\pi r} = \frac{i}{8\pi} \int_{-\infty}^{\infty} dk_\rho \frac{k_\rho}{k_z} H_0^{(1)}(k_\rho \rho) \exp(ik_z |z|) \tag{39}$$

and the Bessel equation for $H_0^{(1)}$

$$\frac{1}{\rho} \frac{\partial}{\partial \rho} \rho \frac{\partial}{\partial \rho} H_0^{(1)} + k_\rho^2 H_0^{(1)} = 0 \tag{40}$$

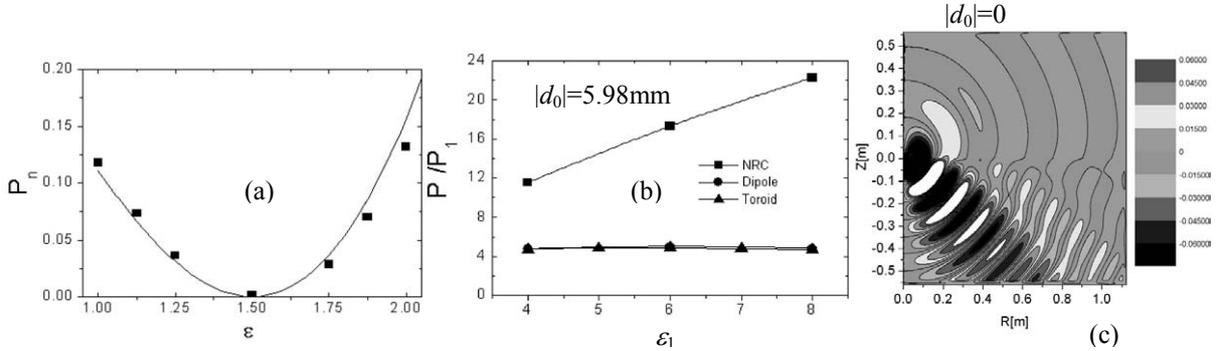

**Figure 6.** Radiation properties of a combination of a first-order toroidal solenoid ($R$=0.5cm, $d$=1cm) and an electric dipole ($L_d$=0.9cm) at 1GHz. Dependence of the normalized emitted power $P_n = 12\pi c P / \mu_0 (\omega I_d L_d)^2$ on the relative dielectric permittivity in a homogeneous ambient material in (a). The value of $\widetilde{\varepsilon}$ is $\widetilde{\varepsilon} = 1.5$. The solid curve is $\sqrt{\varepsilon}(1 - \varepsilon/\widetilde{\varepsilon})^2$ and the squares are the numerical result. The ratio between the powers $P$ and $P_1$ emitted in the materials with dielectric permittivity $\varepsilon$ and $\varepsilon_1$, respectively, as a function of $\varepsilon_1$ with $\varepsilon = \widetilde{\varepsilon} = 1$ in (b). Triangles, circles and squares refer to a radiating toriod, dipole, and a combination of a toroid and a dipole. Distribution of the azimuthal magnetic field component $H_\varphi$ for a combination of toroidal solenoid ($R$=2cm, $d$=4cm) and an electric dipole ($L_d$=1.8cm) in (c). The interface coincides with the equatorial plane of the toroid. The other parameters are $\varepsilon = \widetilde{\varepsilon} = 1$ and $\varepsilon_1$ =4.

(38) can be rewritten as

$$H_z = \frac{ik_0^2 T_2}{8\pi} \int_{-\infty}^{\infty} dk_\rho \frac{k_\rho^3}{k_z} H_0^{(1)}(k_\rho \rho) \exp(ik_z |z|). \tag{41}$$

From (41) it is clear that the source function in (37) is

$$H_S(k_\rho) = H_m(k_\rho) + H_{T2}(k_\rho) = \frac{ik_\rho^3}{8\pi k_z} (k_0^2 T_2 + m). \tag{42}$$

Equation (42) shows that the radiation patterns of an infinitesimal supertoroid and a magnetic dipole are identical in the presence of an interface. In other words, if the compensation condition $k_0^2 T_2 + m = 0$, obtained in a homogeneous material with dielectric permittivity $\varepsilon$ from (30) is satisfied, the presence of the interface will not cause the system to radiate. In the presence of the interface the second-order supertoroid is equivalent to a magnetic dipole with the magnitude of its dipole moment given by $|k_0^2 T_2|$.

In [15] the emission properties of a first-order toroidal solenoid, electric dipole and their combination have been studied and compared. As Fig. 6a shows, although the size of both the toroid and the dipole is finite, reasonable agreement with the analytical result (27) has been obtained at 1GHz in a homogeneous medium. Fig 6b shows that the radiation patterns of the electric dipole and the first order toroidal solenoid are identical in the presence of an interface, as (42) implies. Their combination, however, still emits power while according to (42) it should not. The residual radiation can be attributed to the leading-order uncompensated multipole moment (not taken into account in (42)) of the structure. This multipole is unimportant when the toroid and the dipole are acting alone, because its magnitude is much lower than that of the dipole moments. However, when the net dipole moment of the system is zero (the compensation condition $\varepsilon = \tilde{\varepsilon}$ is satisfied) the next-order multipole moment of the system becomes important. Indeed, the radiation pattern of a combination of a toroid and dipole is not dipole-like, as Fig 6b suggests. Its dependence on the dielectric perimittivity of the substrate material is also much stronger compared to that of a toroid or a dipole acting alone. Fig. 6c shows the distribution of the magnetic field for a combination of a first-order toroidal solenoid and an electric dipole. In this case the interface between the two media coincides with the equatorial plane of the toroid. Because neither of the two media contains the entire system, the same line of considerations, which led us to (42), is not applicable to this case. The problem nevertheless can be solved numerically.

## 7. NONRADIATING CONFIGURATIONS DRIVEN BY LEFT-HANDED METAMATERIALS

The nonradiaitng systems are, in fact, sources that do not produce fields outside their own volume [6, 11]. The non-propagating fields that exist inside such a source might be useful for applications such as sensors – disturbing the configuration of these fields will generate radiation. It is clear however that systems capable of generating non-propagating fields inside an arbitrary large (on a wavelength scale) volumes are needed. The combinations of toroids and dipoles discussed so far do not posses this property. They only work in the limit $d << \lambda$.

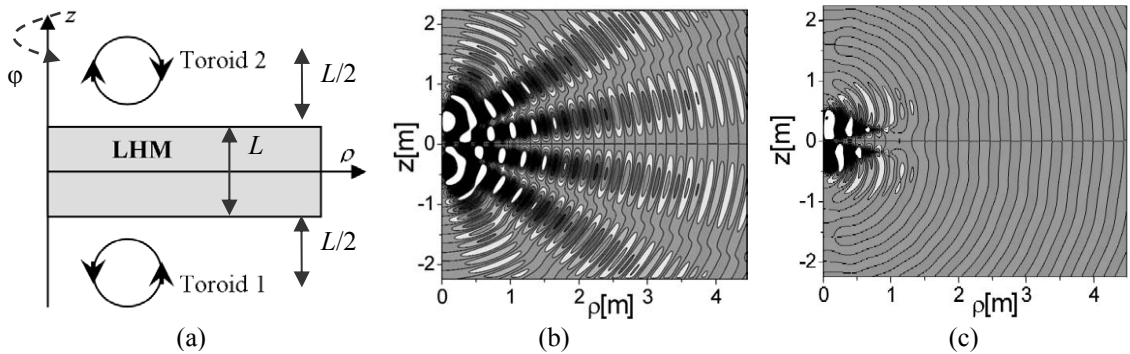

**Figure 7.** Two identical first-order toroids driven by π-out-of-phase currents at 1GHz in the focal planes of a "perfect" lens – a disc made of a left-handed metamaterial (a). Gray-scale map of the azimuthal magnetic field component created by the toroids in the absence of the lens is shown in (b). The corresponding result in the presence of the lens – (c). The currents in (b) and (c) are identical but the power emitted by the system in the absence of the lens is about 20 times higher. Identical scale is used to produce the graphs (b) and (c). White and black colors mean large positive and negative values of the field, respectively, while regions where the absolute value of the field is close to zero are represented with grey tones. The ratio between the collision frequency in the Drude model and the excitation frequency is 0.01. The lens radius is four wavelengths and the distance between the two emitters – 2.6 wavelengths.

Fig. 7a shows a couple of toroidal solenoids driven by π-out-of-phase currents in the focal planes of a disc made of a left-handed metamaterial (LHM). It has been shown (see e.g. [20-25]) that a slab of such a material has the properties of an aberrations-free lens, capable of providing resolution below the diffraction limit. It is easy to show that the interaction of the image of the first emitter with the second emitter and vice-versa limits the power emitted by the system. As a result of this interaction, the electromagnetic field remains confined between the two emitters and the LHM disc. To verify this, we have studied numerically the system depicted on Fig. 7a. The Drude model is used for both the permittivity and permeability of the lens. Details on FDTD modeling of left-handed metamaterials are given in [22, 23].

Fig. 7 shows a comparison between the fields generated by the same current system (two toroidal solenoids): in the presence of the lens (c) and in its absence (b). As this comparison shows, the lens limits strongly the amount of power radiated by the system: by a factor of 20 in the case presented in Fig. 7. Strong non-propagating fields are generated in the regions between each of the emitters and the lens. The amount of residual emitted power strongly depends of the parameter values such as collision frequency, size of the lens, distance between the emitters, etc. The fact that one and same current distribution radiates different amounts of power (Fig 7b and Fig. 7c) is not a contradiction with the energy conservation law. It can be shown that the power input from the generator, driving the currents in the two emitters, always equals the sum of the power dissipated in the volume of the lens and the radiated power. By properly choosing the system parameters the latter can be minimized. The details will be given in a separate paper. It should be noted however that the toroids are not the only type of emitters that can work in the nonradiating system of Fig. 7a. Qualitatively the same results are expected with any two identical radiators.

In conclusion nonradiating systems based on toroidal and supertoroidal currents are studied. The general criteria for absence of radiation known from the literature are applied to these systems and a simple connection is established between the current density spectrum and the uncompensated electromagnetic potentials. It is shown that the radiation patterns of an infinitesimal supertoroid and a dipole are identical in the presence of an inhomogeneity in the form of a planar interface between two dielectric media. Design of a new type of non-radiating system based on left-handed metamaterial is suggested.

## ACKNOWLEDGMENTS

This work is supported by the Engineering and Physical Sciences Research Council (UK) under the Adventure Fund Program.

## REFERENCES

1. G. H. Goedecke, *Phys. Rev.* **135**, pp. B281-B288, 1964.
2. D. Bohm and M. Weinstein, *Phys. Rev.* **74**, pp.1789-1798, 1948.
3. A. J. Devaney and G. C. Sherman, *IEEE Trans. on Antennas and Propagation* **AP-30**, pp.1034-1037, 1982.
4. E. A. Marengo and R. W. Ziolkowski, *J. Opt. A: Pure Appl. Opt.* **2**, pp.179-187, 2000.
5. E. A. Marengo and R. W. Ziolkowski, *IEEE Trans. on Antennas and Propagation* **48**, pp.1553-1562, 2000.
6. N. K. Nikolova and Y. S. Rickard, *Phys. Rev.* **E**, **71**, 016617, 2005.
7. G. N. Afanasiev and V. M. Dubovik, *Phys. Part. Nuclei* **29**, pp. 366-391, 1998.
8. G. N. Afanasiev and Yu. P. Stepanovsky, *J. Phys. A: Math. Gen.* **28**, pp. 4565-4580, 1995.
9. G. N. Afanasiev, *J. Phys. D: Appl. Phys.* **34**, pp. 539-559, 2001.
10. K. Kim and E. Wolf, *Opt. Commun.* **59**, pp. 1-6, 1986.
11. E. A. Marengo and A. J. Devaney, *Phys. Rev. E* **70**, 037601, 2004.
12. A. J. Devaney and E. Wolf, *Phys. Rev. D* **8**, pp.1044-1047, 1973.
13. J. D. Jackson, *Classical Electrodynamics*, Wiley, 1999.
14. V. M. Dubovik and V. V. Tugushev, *Phys. Reports* **187**, pp. 145-202, 1990.
15. A. D. Boardman, K. Marinov, N. Zheludev and V. A. Fedotov, *Phys. Rev. E*, 2005, accepted for publication.
16. A. Taflove and S. Hagness, *Computational electrodynamics: the finite-difference time-domain method*, Artech House (2000).
17. J. A. Stratton, *Electromagnetic Theory*, Mc-Graw Hill, 1941.
18. A. Banos, *Dipole Radiation in the Presence of a Conducting Half-Space*, Pergamon Press, 1966.
19. J. A. Kong, *Electromagnetic Wave Theory*, John Wiley, 1986.
20. R. A. Silin, *Opt. Spectrosc.* **44**, 109, 1978.
21. J. B. Pendry, *Phys. Rev. Lett.* **85**, pp. 3966-3969, 2000.
22. R. Ziolkowski, *Opt. Express* **11**, pp 662-681, 2003.
23. R. Ziolkowski and E. Heyman, *Phys. Rev. E* **64**, 056625, 2001.
24. X. S. Rao and C. K. Ong, *Phys. Rev. E.* **68**, 067601, 2003.
25. L. Chen, S. He and L. Shen, *Phys. Rev. Lett.* **92**, 107404, 2004.